\documentclass[journal]{IEEEtran}

\ifCLASSINFOpdf
\usepackage[pdftex]{graphicx}
\usepackage[caption=false, font=normalsize, labelfont=sf, textfont=sf]{subfig}
\graphicspath{{../pdf/}{../jpg/}}
\else
  \usepackage[dvips]{graphicx}
  \graphicspath{{../eps/}}
  \DeclareGraphicsExtensions{.eps}
\fi

\usepackage{cite}
\usepackage[cmex10]{amsmath}
\usepackage{amssymb}
\usepackage{amsmath}
\usepackage{fixmath}

\usepackage{enumitem}
\usepackage[short]{optidef}
\usepackage{comment}
\usepackage{color}
\usepackage{soul}

\usepackage{tikz}
\usepackage{pgfplots}
\pgfplotsset{compat=newest}

\usepackage{hyperref}
\usepackage[capitalise]{cleveref}
\Crefname{equation}{}{}

\begin{document}
\title{Reliability-Aware Probabilistic Reserve Procurement}

\author{Lars~Herre,~\IEEEmembership{Member,~IEEE,}
Pierre~Pinson,~\IEEEmembership{Fellow,~IEEE,}
Spyros~Chatzivasileiadis,~\IEEEmembership{Senior~Member,~IEEE,}
    \thanks{L. Herre and S. Chatzivasileiadis are with the Center for Electric Engineering, DTU Denmark Technical University, 2800 Kgs. Lyngby, Denmark. e-mail: \{lfihe,spchatz\}@elektro.dtu.dk. P. Pinson is with the Department of Technology, Management and Economics, DTU Denmark Technical University, 2800 Kgs. Lyngby, Denmark. e-mail: \{ppin\}@dtu.dk.}
}% <-this % stops a space

\maketitle

\begin{abstract}
%Probabilistic approaches have been proposed for reserve sizing and allocation and here, we propose reliability-aware reserve procurement.
Current reserve procurement approaches ignore the stochastic nature of reserve asset availability itself and thus limit the type and volume of reserve offers.
This paper develops a reliability-aware probabilistic approach that allows renewable generators to offer reserve capacity with reliability attributes.
Offers with low reliability are priced at lower levels.
The original non-convex market clearing problem is approximated by a MILP reformulation.
The proposed probabilistic reserve procurement allows restricted reserve providers to enter the market, thereby increases liquidity and has the potential to lower procurement costs in power systems with high shares of variable renewable energy sources. 
\end{abstract}

\begin{IEEEkeywords}
electricity markets, reserve procurement, risk management, power system operation.
\end{IEEEkeywords}

\section{Introduction}
%Reference \cite{Wang2020} 

Reliable power system operation requires procurement of sufficient reserve capacity to account for unplanned `credible' contingencies.
Current approaches determine these requirements using deterministic security margins which aim to ensure a prespecificied probabilistic reliability index, such as EENS, LOLP, SAIDI, SAIFI, etc. \cite{EC2016}. 
Originally, however, these indexes are computed in expectation and extracted from probability density functions which contain the full set of information. As a result, existing methods are unable to trade-off the risk of potential contingency and its associated volume against the reliability of a procured reserve and its associated volume.

%%%% For Pierre to review and add literature %%%%
Energy markets with probabilistic offers have been investigated in \cite{Papakonstantinou2016}. 
Reference \cite{Zhao2015} analysed aggregation problems and risky power markets.
Chance constrained programming for joint clearing of energy and reserves in peer to peer markets has been applied in \cite{Guo2021}.
There exist several papers on the joint clearing of energy and reserves under uncertainty, where reliability awareness is implicitly included in the stochastic formulation, e.g., \cite{Ordoudis2021} using distributionally robust optimization. 
Here, we focus on the reserve clearing problem with the aim of developing a more tractable market clearing tool for system operators (SO) which can easily be incorporated in existing deterministic frameworks.

The literature is rich in probabilistic methods for SOs to determine the \emph{reserve requirement}, e.g., \cite{Doherty2005,Matos2011,DeVos2013, DeVos2019, Mieth2021}. 
A methodology which quantifies the reserve need taking into account the uncertain nature of wind power is presented in \cite{Doherty2005}. System reliability is used as an objective measure to determine the effect of increasing wind power penetration. Reference \cite{Matos2011} proposes a reserve management tool to support the SO in defining the operating reserve needs.
An overview of probabilistic sizing methods is given in \cite{DeVos2013}, and dynamic reserve sizing is investigated in, e.g., \cite{DeVos2019} as a function of the system risk. Varying renewable energy sources (RES) are commonly viewed as the reason for increased reserve requirements. Reference \cite{Mieth2021} proposes to account for risk-aware reserve dimensioning, allocation and deliverability in a security constrained unit commitment framework by learning risk-aware reserve activation factors. The proposed methods would result in increased liquidity in reliability-aware markets.

The reviewed references describe a methodologies to support SOs in defining the \emph{reserve requirement}, due to increasing uncertain renewable generation.
However, none of the previous works have addressed the probabilistic selection of the type and amount of reserve in the \emph{procurement} stage based on its individual reliability.
Instead of accounting for uncertainty in the reserve sizing, here, we present a market framework to transparently include uncertainty as a reserve offer attribute.
Probabilistic procurement would allow the reserve provider to specify the offer reliability, i.e., the probability of reserve availability when activated in real-time.
%In other words, the reliability of reserve delivery when needed in real-time.

Reserves from conventional generators are viewed to have a reliability of 100\,\% when ignoring unplanned events (\emph{force majeure}). Reserves from renewable energy sources (RES), battery energy storage (BES), and demand response providers (DR), however, can only guarantee a small percentage of their predicted available capacity with 100\,\% reliability of delivery. This is due to different sources of uncertainty, prediction errors, and variability. If the reliability requirement is lowered to less than 100\,\%, however, these \emph{restricted reserves providers} (RRPs) can commit more capacity to SOs, at significantly lower prices. 
For instance, a provider that knows their reserve will only be available with 90\,\% probability will not bid in the market, since they will be heavily penalised if an instance occurs where they cannot deliver. But if we are able to embrace these uncertainties in a market clearing scheme, a water heater, for example, may be willing to offer down-reserves at half the price, if given the chance to not deliver it 1 out of 10 times. This not only increases the liquidity in the market, but it can also reduce prices, as
it can be argued that the relation between reliability and offered prices is non-linear for RRPs, with price offers decreasing exponentially for each percentage point of decreased reliability.

%Contingency analyses and risk-based calculations for power system short-term planning are typically modelled with security-constrained unit commitment problem (SCUC) \cite{Mieth2021}. 
In this paper, we introduce the concept of our proposed reliability-aware reserve clearing. In this first stage, we ignore the network; the proposed method can be applied to a reserve zone where intra-zonal constraints are commonly ignored \cite{Mieth2021}.
Specifically, our contributions are the following:
\begin{itemize}
    \item We propose a reliability-aware reserve market clearing tool which is (i) simple to use for reserve providers and (ii) tractable and reliable for SOs that clear the market. 
    \item We challenge the conventional idea of 100\,\% reserve availability, which no reserve provider can guarantee if \emph{force majeure} is internalised by providers. Rethinking this availability metric opens the market for new players, increases competition and thus liquidity, and lowers cost.
\end{itemize}
Opposite to stochastic energy market clearing in, e.g., \cite{Papakonstantinou2016}, here, the volume at equilibrium is not determined by social welfare maximization, but by system operation cost minimization with reliability attributes of bids.
%%%%%%%%%%%%%%%%%%%%%%%%%%
% Opposite to stochastic energy market clearing, here, the volume at equilibrium is not determined by social welfare maximization, but by system risk and system operation cost minimization, which includes parameters outside of the market clearing.

%%%%%%%%%%%%%%%%%%%%%%%%%%

The rest of this paper is structured as follows.
\cref{sec:Scene} introduces the conceptual idea, provides a motivating example and lists benefits of such a market. 
\cref{sec:marketframework} introduces the market framework including actors and timeline, and gives the mathematical problem formulation. \cref{sec:problem} presents an exact problem reformulation, and a linear approximation. A small tangible case study and a large national level case study are investigated in \cref{sec:Case}. \cref{sec:Disc} discusses the results and possible extensions. We conclude with key take-aways in \cref{sec:Conc}.

\section{Setting the Scene} \label{sec:Scene}
To better explain the motivation behind the method detailed in the following, this section provides a simple example and lists benefits of such a reliability-aware reserve market.

\subsection{Motivating Example}
To illustrate the proposed market framework, we offer a simple example. A set of reserve offers with its characteristics is sorted by reliability in \cref{tab:Reliab}. %, which is defined as the probability of availability.
\begin{table}[t]
  \centering
  \caption{Simple Example of Reliability-Aware Reserve Offers}
    %\resizebox{0.65\columnwidth}{!}{%
    \setlength{\tabcolsep}{4pt}
    \begin{tabular}{c rrr}
    \hline
    Offer  & Volume     & Cost             & Reliability \\
    \hline
    1       & 100 MW    &100 \textdollar/MW & 99 \%  \\
    2       & 100 MW    & 55 \textdollar/MW & 98 \%  \\
    3       & 100 MW    & 40 \textdollar/MW & 95 \%  \\
    4       & 100 MW    & 25 \textdollar/MW & 90 \%  \\
    5       & 100 MW    & 11 \textdollar/MW & 70 \%  \\
    6       & 100 MW    & 10 \textdollar/MW & 70 \%  \\
    \hline 
    \end{tabular} %}
  \label{tab:Reliab} 
\vspace{-0.2cm}
\end{table}
Let the reserve requirement be 100\,MW with 99\,\% reliability.
%Reserve procurement can be conducted with three different strategies: 
Reserve 1 alone would satisfy the required reserve volume for 100\,\textdollar. 

However, the combined procurement of reserves 2 and 3 can offer similar or higher reliability and cost less. 
Assuming that the uncertainty from different offers are uncorrelated (independence assumption), we can calculate the \emph{joint availability} by simple multiplication.
The probability of \emph{joint \textbf{un}availability} (0\,MW) of both reserves 2 and 3 is only 0.1\,\%, while the probability of \emph{joint availability} (200 MW) is 93.1\,\%. Thus, a volume of at least 100 MW is available with 99.9\,\% reliability, which is higher than the most reliable reserve itself. The combined cost for procuring reserves 2 and 3 is 95\,\textdollar.
Similarly, reserves 4, 5 and 6 have a joint reliability of 99.1\,\% with a cost of only 46\,\textdollar. This illustrates the lowered procurement cost due to higher volumes that are procured from RRPs with lower reliability. 

Notably, this problem becomes more challenging when offer volumes are not uniform. The full mathematical description of reliability-aware reserve clearing is set out in \cref{sec:marketframework}.

\subsection{Benefits for Restricted Reserve Providers (RRPs)}
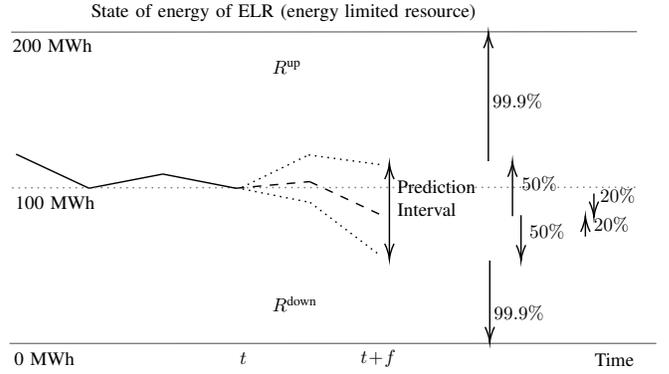
\begin{figure}[t]
\centering
\resizebox{\columnwidth}{!}{\tikzset{every picture/.style={line width=0.75pt}} %set default line width to 0.75pt        

\begin{tikzpicture}[x=0.75pt,y=0.75pt,yscale=-1,xscale=1]
%uncomment if require: \path (0,278); %set diagram left start at 0, and has height of 278

%Straight Lines [id:da41355886174326173] 
\draw [color={rgb, 255:red, 128; green, 128; blue, 128 }  ,draw opacity=1 ]   (4.33,245.07) -- (459.75,245.05) ;
%Straight Lines [id:da2670150611727673] 
\draw [color={rgb, 255:red, 128; green, 128; blue, 128 }  ,draw opacity=1 ]   (5.49,25.7) -- (446.1,25.31) ;
%Straight Lines [id:da5882499535788928] 
\draw    (341.5,116.63) -- (341.5,28) ;
\draw [shift={(341.5,26)}, rotate = 450] [color={rgb, 255:red, 0; green, 0; blue, 0 }  ][line width=0.75]    (10.93,-3.29) .. controls (6.95,-1.4) and (3.31,-0.3) .. (0,0) .. controls (3.31,0.3) and (6.95,1.4) .. (10.93,3.29)   ;
%Straight Lines [id:da4625791901601588] 
\draw    (60.33,135.67) -- (112.33,125.67) ;
%Straight Lines [id:da9640492001252021] 
\draw [color={rgb, 255:red, 128; green, 128; blue, 128 }  ,draw opacity=1 ] [dash pattern={on 0.84pt off 2.51pt}]  (6.15,135.37) -- (446.77,134.98) ;
%Straight Lines [id:da6230058605463198] 
\draw    (112.33,125.67) -- (163.67,135.67) ;
%Straight Lines [id:da535633987508588] 
\draw  [dash pattern={on 4.5pt off 4.5pt}]  (215.67,131) -- (267,155) ;
%Straight Lines [id:da5978736906852877] 
\draw    (9,111.67) -- (60.33,135.67) ;
%Straight Lines [id:da717451492674819] 
\draw  [dash pattern={on 4.5pt off 4.5pt}]  (163.67,135.67) -- (215.67,131) ;
%Straight Lines [id:da3471290719864968] 
\draw    (342.5,242.35) -- (342.5,186.63) ;
\draw [shift={(342.5,244.35)}, rotate = 270] [color={rgb, 255:red, 0; green, 0; blue, 0 }  ][line width=0.75]    (10.93,-3.29) .. controls (6.95,-1.4) and (3.31,-0.3) .. (0,0) .. controls (3.31,0.3) and (6.95,1.4) .. (10.93,3.29)   ;
%Straight Lines [id:da009253613914891101] 
\draw  [dash pattern={on 0.84pt off 2.51pt}]  (216,145.67) -- (265.33,183) ;
%Straight Lines [id:da2788045307334519] 
\draw  [dash pattern={on 0.84pt off 2.51pt}]  (168.41,136.2) -- (216,145.67) ;
%Straight Lines [id:da2876140067397215] 
\draw  [dash pattern={on 0.84pt off 2.51pt}]  (214.93,112.19) -- (265.33,119) ;
%Straight Lines [id:da49669364473444255] 
\draw  [dash pattern={on 0.84pt off 2.51pt}]  (167.99,135.04) -- (214.93,112.19) ;
%Straight Lines [id:da8904134214921209] 
\draw    (271.83,183.67) -- (271.83,120.33) ;
\draw [shift={(271.83,118.33)}, rotate = 450] [color={rgb, 255:red, 0; green, 0; blue, 0 }  ][line width=0.75]    (10.93,-3.29) .. controls (6.95,-1.4) and (3.31,-0.3) .. (0,0) .. controls (3.31,0.3) and (6.95,1.4) .. (10.93,3.29)   ;
\draw [shift={(271.83,185.67)}, rotate = 270] [color={rgb, 255:red, 0; green, 0; blue, 0 }  ][line width=0.75]    (10.93,-3.29) .. controls (6.95,-1.4) and (3.31,-0.3) .. (0,0) .. controls (3.31,0.3) and (6.95,1.4) .. (10.93,3.29)   ;
%Straight Lines [id:da17570761979095484] 
\draw    (358.5,155.33) -- (358.5,118.67) ;
\draw [shift={(358.5,116.67)}, rotate = 450] [color={rgb, 255:red, 0; green, 0; blue, 0 }  ][line width=0.75]    (10.93,-3.29) .. controls (6.95,-1.4) and (3.31,-0.3) .. (0,0) .. controls (3.31,0.3) and (6.95,1.4) .. (10.93,3.29)   ;
%Straight Lines [id:da7131289195333665] 
\draw    (364.5,154.67) -- (364.5,184.67) ;
\draw [shift={(364.5,186.67)}, rotate = 270] [color={rgb, 255:red, 0; green, 0; blue, 0 }  ][line width=0.75]    (10.93,-3.29) .. controls (6.95,-1.4) and (3.31,-0.3) .. (0,0) .. controls (3.31,0.3) and (6.95,1.4) .. (10.93,3.29)   ;
%Straight Lines [id:da6528255224409678] 
\draw    (409.83,169.67) -- (409.83,157.43) ;
\draw [shift={(409.83,155.43)}, rotate = 450] [color={rgb, 255:red, 0; green, 0; blue, 0 }  ][line width=0.75]    (10.93,-3.29) .. controls (6.95,-1.4) and (3.31,-0.3) .. (0,0) .. controls (3.31,0.3) and (6.95,1.4) .. (10.93,3.29)   ;
%Straight Lines [id:da4271222046998655] 
\draw    (415.83,139.43) -- (415.83,154) ;
\draw [shift={(415.83,156)}, rotate = 270] [color={rgb, 255:red, 0; green, 0; blue, 0 }  ][line width=0.75]    (10.93,-3.29) .. controls (6.95,-1.4) and (3.31,-0.3) .. (0,0) .. controls (3.31,0.3) and (6.95,1.4) .. (10.93,3.29)   ;

% Text Node
\draw (36.03,255.75) node   [align=left] {\begin{minipage}[lt]{42.58pt}\setlength\topsep{0pt}
0 MWh
\end{minipage}};
\draw (195,255.75) node   [align=left] {\begin{minipage}[lt]{42.58pt}\setlength\topsep{0pt}
$t$
\end{minipage}};
\draw (280,255.75) node   [align=left] {\begin{minipage}[lt]{42.58pt}\setlength\topsep{0pt}
$t\!+\!f$
\end{minipage}};
% Text Node
\draw (222.41,11.75) node   [align=left] {\begin{minipage}[lt]{240.84pt}\setlength\topsep{0pt}
State of energy of ELR (energy limited resource)
\end{minipage}};
% Text Node
\draw (41.59,146.03) node   [align=left] {\begin{minipage}[lt]{49.61pt}\setlength\topsep{0pt}
100 MWh
\end{minipage}};
% Text Node
\draw (39.74,35.03) node   [align=left] {\begin{minipage}[lt]{49.61pt}\setlength\topsep{0pt}
200 MWh
\end{minipage}};
% Text Node
\draw (310.3,142.91) node   [align=left] {\begin{minipage}[lt]{49.23pt}\setlength\topsep{0pt}
Prediction\\Interval
\end{minipage}};
% Text Node
\draw (343,67.33) node [anchor=north west][inner sep=0.75pt]    {$99.9\%$};
% Text Node
\draw (344,216.33) node [anchor=north west][inner sep=0.75pt]    {$99.9\%$};
% Text Node
\draw (188,44) node [anchor=north west][inner sep=0.75pt]    {$R^\text{up}$};
% Text Node
\draw (188,210) node [anchor=north west][inner sep=0.75pt]    {$R^\text{down}$};
% Text Node
\draw (436.96,256.7) node   [align=left] {\begin{minipage}[lt]{30.88pt}\setlength\topsep{0pt}
Time
\end{minipage}};
% Text Node
\draw (369,159) node [anchor=north west][inner sep=0.75pt]    {$50\%$};
% Text Node
\draw (363.67,125.67) node [anchor=north west][inner sep=0.75pt]    {$50\%$};
% Text Node
\draw (414.33,154.33) node [anchor=north west][inner sep=0.75pt]    {$20\%$};
% Text Node
\draw (418.71,134.14) node [anchor=north west][inner sep=0.75pt]    {$20\%$};

\end{tikzpicture}}
\caption{Illustration of state of energy (SOE) of an energy limited resource with its energy bounds and reserve volume and reliability.}
\label{fig:reliabilityoffers}
\vspace{-0.2cm}
\end{figure}

\cref{fig:reliabilityoffers} aims to illustrate the benefit for RRPs with the example of a generic energy limited resource (ELR), such as an energy storage system or certain types of demand response with storage capability. The upper and lower bounds on the state-of-energy (SOE) limit the available power and energy capacity of the ELR. At a given time $t$, and looking into the future $t+f$, the ELR operator faces uncertainty on their forecast SOE due to participation in other energy or ancillary service markets. With longer prediction horizon $f$, this uncertainty increases, i.e., the prediction interval of the SOE increases. However, the remaining margin between the SOE prediction interval and the SOE bound can be used as a reliable up ($R^\text{up}$) or down-reserve ($R^\text{down}$). 

Instead of claiming 100\,\% reliability, which is current practice in reserve markets, we encourage the more realistic use of, e.g., 99.9\,\% reliability in \cref{fig:reliabilityoffers} which would more transparently include events of \emph{force majeure}. These events are currently not considered in the deterministic market clearing of reserve or energy markets, with the exception of $N-1$ approaches.

Furthermore, the ELR operator may also offer part of their uncertain reserve with lower reliability. For instance, the probability of an SOE realization between the expected mean and the worst-case-minimum is 50\,\%. In other words, the real SOE ends up below the expected SOE with 50\,\% probability. Therefore, the margin between expected SOE and worst-case-maximum SOE can be offered as upward reserve with 50\,\% reliability. 
Alternatively, this margin can be decomposed into smaller power intervals of different reliability levels.

\subsection{Benefits for System Operators (SOs)}
SOs aim to minimise procurement and operation cost. Opening the reserve market to additional players such as RRPs is likely to increase liquidity. Furthermore, under the assumption that offers with lower reliability are priced lower, a reliability-aware market has the potential to lower costs while maintaining the same total level of reliability for the procured reserves, as we show in \cref{sec:Case}.
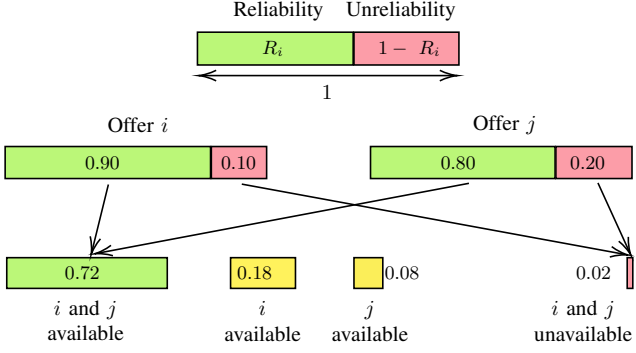
\begin{figure}[t]
\centering
\resizebox{\columnwidth}{!}{\tikzset{every picture/.style={line width=0.75pt}} %set default line width to 0.75pt        

\begin{tikzpicture}[x=0.75pt,y=0.75pt,yscale=-1,xscale=1]
%uncomment if require: \path (0,491); %set diagram left start at 0, and has height of 491

%Shape: Rectangle [id:dp08116221137588275] 
\draw  [fill={rgb, 255:red, 187; green, 247; blue, 120 }  ,fill opacity=1 ] (225.83,139) -- (323,139) -- (323,159.48) -- (225.83,159.48) -- cycle ;
%Shape: Rectangle [id:dp590582531182174] 
\draw  [fill={rgb, 255:red, 252; green, 157; blue, 168 }  ,fill opacity=1 ] (323.5,139) -- (389,139) -- (389,159.48) -- (323.5,159.48) -- cycle ;
%Shape: Rectangle [id:dp8343516326023293] 
\draw  [fill={rgb, 255:red, 187; green, 247; blue, 120 }  ,fill opacity=1 ] (105.83,210.4) -- (234,210.4) -- (234,230.4) -- (105.83,230.4) -- cycle ;
%Shape: Rectangle [id:dp7049963687619323] 
\draw  [fill={rgb, 255:red, 252; green, 157; blue, 168 }  ,fill opacity=1 ] (234.5,210.4) -- (269,210.4) -- (269,230.4) -- (234.5,230.4) -- cycle ;
%Straight Lines [id:da10418744592997498] 
\draw    (228.58,166.1) -- (386,166.1) ;
\draw [shift={(388,166.1)}, rotate = 180] [color={rgb, 255:red, 0; green, 0; blue, 0 }  ][line width=0.75]    (10.93,-3.29) .. controls (6.95,-1.4) and (3.31,-0.3) .. (0,0) .. controls (3.31,0.3) and (6.95,1.4) .. (10.93,3.29)   ;
\draw [shift={(226.58,166.1)}, rotate = 0] [color={rgb, 255:red, 0; green, 0; blue, 0 }  ][line width=0.75]    (10.93,-3.29) .. controls (6.95,-1.4) and (3.31,-0.3) .. (0,0) .. controls (3.31,0.3) and (6.95,1.4) .. (10.93,3.29)   ;
%Shape: Rectangle [id:dp13406939790672534] 
\draw  [fill={rgb, 255:red, 187; green, 247; blue, 120 }  ,fill opacity=1 ] (333.83,210.4) -- (449,210.4) -- (449,230.4) -- (333.83,230.4) -- cycle ;
%Shape: Rectangle [id:dp6283757622913291] 
\draw  [fill={rgb, 255:red, 252; green, 157; blue, 168 }  ,fill opacity=1 ] (449.5,210.4) -- (497,210.4) -- (497,230.4) -- (449.5,230.4) -- cycle ;
%Shape: Rectangle [id:dp6560720242200817] 
\draw  [fill={rgb, 255:red, 187; green, 247; blue, 120 }  ,fill opacity=1 ] (106.83,279.82) -- (207,279.82) -- (207,299) -- (106.83,299) -- cycle ;
%Shape: Rectangle [id:dp31494784199630876] 
\draw  [fill={rgb, 255:red, 252; green, 157; blue, 168 }  ,fill opacity=1 ] (494,279.82) -- (497.5,279.82) -- (497.5,299) -- (494,299) -- cycle ;
%Shape: Rectangle [id:dp12021227918268962] 
\draw  [fill={rgb, 255:red, 255; green, 242; blue, 92 }  ,fill opacity=1 ] (246.5,279.82) -- (287,279.82) -- (287,299) -- (246.5,299) -- cycle ;
%Shape: Rectangle [id:dp0022909492813716348] 
\draw  [fill={rgb, 255:red, 251; green, 238; blue, 89 }  ,fill opacity=1 ] (323.5,279.82) -- (341.5,279.82) -- (341.5,299) -- (323.5,299) -- cycle ;
%Straight Lines [id:da4791570868830295] 
\draw    (253.5,233.88) -- (492.53,277.76) ;
\draw [shift={(494.5,278.13)}, rotate = 190.4] [color={rgb, 255:red, 0; green, 0; blue, 0 }  ][line width=0.75]    (10.93,-3.29) .. controls (6.95,-1.4) and (3.31,-0.3) .. (0,0) .. controls (3.31,0.3) and (6.95,1.4) .. (10.93,3.29)   ;
%Straight Lines [id:da21828727104959733] 
\draw    (475.75,233.13) -- (495.42,276.3) ;
\draw [shift={(496.25,278.13)}, rotate = 245.51] [color={rgb, 255:red, 0; green, 0; blue, 0 }  ][line width=0.75]    (10.93,-3.29) .. controls (6.95,-1.4) and (3.31,-0.3) .. (0,0) .. controls (3.31,0.3) and (6.95,1.4) .. (10.93,3.29)   ;
%Straight Lines [id:da229356992724572] 
\draw    (395.25,233.38) -- (163.72,277) ;
\draw [shift={(161.75,277.38)}, rotate = 349.33000000000004] [color={rgb, 255:red, 0; green, 0; blue, 0 }  ][line width=0.75]    (10.93,-3.29) .. controls (6.95,-1.4) and (3.31,-0.3) .. (0,0) .. controls (3.31,0.3) and (6.95,1.4) .. (10.93,3.29)   ;
%Straight Lines [id:da5670066253138653] 
\draw    (170.25,234.13) -- (159.98,275.93) ;
\draw [shift={(159.5,277.88)}, rotate = 283.8] [color={rgb, 255:red, 0; green, 0; blue, 0 }  ][line width=0.75]    (10.93,-3.29) .. controls (6.95,-1.4) and (3.31,-0.3) .. (0,0) .. controls (3.31,0.3) and (6.95,1.4) .. (10.93,3.29)   ;

% Text Node
\draw (335.53,125.46) node   [align=left] {\begin{minipage}[lt]{130.83pt}\setlength\topsep{0pt}
Reliability \ \ Unreliability
\end{minipage}};
% Text Node
\draw (196.5,196.74) node   [align=left] {\begin{minipage}[lt]{39.89pt}\setlength\topsep{0pt}
Offer $i$
\end{minipage}};
% Text Node
\draw (264.67,143) node [anchor=north west][inner sep=0.75pt]  [font=\small]  {$R_{i} \ \ \ \ \ \ \ \ \ \ \ \ \ \ 1-\ R_{i}$};
% Text Node
\draw (154.17,215) node [anchor=north west][inner sep=0.75pt]  [font=\small]  {$0.90\ \ \ \ \ \ \ \ \ \ \ \ \ \ \ 0.10$};
% Text Node
\draw (311.58,176.19) node   [align=left] {\begin{minipage}[lt]{12.13pt}\setlength\topsep{0pt}
$1$
\end{minipage}};
% Text Node
\draw (424.5,196.74) node   [align=left] {\begin{minipage}[lt]{39.89pt}\setlength\topsep{0pt}
Offer $j$
\end{minipage}};
% Text Node
\draw (376.17,215) node [anchor=north west][inner sep=0.75pt]  [font=\small]  {$0.80\ \ \ \ \ \ \ \ \ \ \ \ \ \ 0.20$};
% Text Node
\draw (155.92,318.91) node   [align=left] {\begin{minipage}[lt]{61.31pt}\setlength\topsep{0pt}
\begin{center}
$i$ and $j$\\available
\end{center}

\end{minipage}};
% Text Node
\draw (141.67,284) node [anchor=north west][inner sep=0.75pt]  [font=\small]  {$0.72$};
% Text Node
\draw (460.67,284) node [anchor=north west][inner sep=0.75pt]  [font=\small]  {$0.02$};
% Text Node
\draw (249.17,284) node [anchor=north west][inner sep=0.75pt]  [font=\small]  {$0.18\ $};
% Text Node
\draw (266.42,319.41) node   [align=left] {\begin{minipage}[lt]{61.31pt}\setlength\topsep{0pt}
\begin{center}
$i$\\available
\end{center}

\end{minipage}};
% Text Node
\draw (333.42,319.41) node   [align=left] {\begin{minipage}[lt]{61.31pt}\setlength\topsep{0pt}
\begin{center}
$j$\\available
\end{center}

\end{minipage}};
% Text Node
\draw (466.42,319.41) node   [align=left] {\begin{minipage}[lt]{61.31pt}\setlength\topsep{0pt}
\begin{center}
$i$ and $j$\\unavailable
\end{center}

\end{minipage}};
% Text Node
\draw (341.17,284) node [anchor=north west][inner sep=0.75pt]  [font=\small]  {$0.08\ $};

\end{tikzpicture}}
\caption{Illustration of the reliability gain from procuring multiple reserves in parallel. The \emph{total} reliability is $0.98$.}
\label{fig:SimpleIntro}
\vspace{-0.2cm}
\end{figure}
A simple example of parallel procurement of two low-priced offers $i$ and $j$ with the same volume $q_i=q_j$ is depicted in \cref{fig:SimpleIntro}. Combining two offers of the same volume that can each deliver their energy with 80\,\% and 90\,\% reliability results in being able to offer a total volume $q=q_i=q_j$ with a reliability of 98\,\%. If we were to consider, however, all bids individually, as conventional markets do, the joint availability of the total volume $q_i+q_j$ is only 72\,\%; in that case, both of these offers would have never entered a conventional market.

%\newpage
\section{Market Framework} \label{sec:marketframework}
This section details the market actors, timeline, probabilistic foundations, and the resulting market clearing problem formulation.
\subsection{Market Actors \& Timeline}
We envision a reliability-aware probabilistic reserve market that is centrally organised by the SO. In practice, we refer to a transmission system operator (in e.g. Europe) or independent system operator (in e.g. North America).
The SO commonly clears the market on day-ahead, hour-ahead, or even minute-ahead basis, and therefore the solution time of the algorithm may become a vital issue.

Offers are submitted by restricted reserve providers (RRP). In reality, even the most reliable provider cannot guarantee 100\,\% reliability due to unplanned events (\emph{force majeure}). While, today, this is only considered as out-of-market tail risk, here, we assume that all market participants are indeed RRPs.

Reliability-aware reserve offer $i$ includes the reserve volume $V_i$, price $P_i$, and reliability $R_i$. 
We assume that the reliability of offers is independent. This assumption is thoroughly discussed in \cref{sec:Disc}.
To clear the reserve market offers can be combined in different ways in order to achieve the reserve volume $Q^s$ with reliability $\Phi^s$ required by the SO. 

\begin{figure}[t]
\centering
\resizebox{\columnwidth}{!}{\input{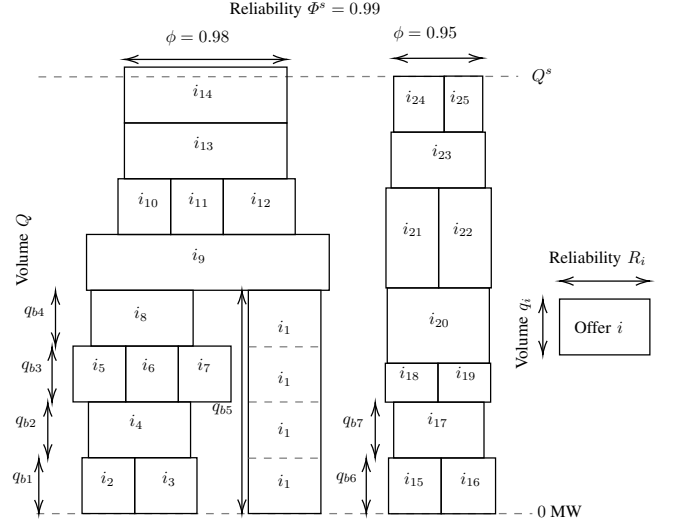}}
\caption{Illustration of volume and reliability stacking. To obtain the target reliability of 99\,\% we build two stacks, one with 98\,\% and the other with 95\,\% reliability. To build each stack, we combine offers horizontally to increase their reliability according to \eqref{eq:relstacking2}, and stack them vertically to increase the procured volume according to \eqref{eq:volstacking1}. Please note that vertical stacking decreases the total reliability according to \eqref{eq:volstacking2}, so each procurement row must have a higher reliability than the target reliability of each stack.}
\label{fig:offerstacking}
\vspace{-0.2cm}
\end{figure}

A possible combination of offers is visualised in \cref{fig:offerstacking}. 
Offers are depicted as rectangles where the height corresponds to their volume and the width corresponds to their reliability. 
The horizontal stacking of offers increases reliability for the same volume. For instance, the horizontal combination of 80\,\% and 90\,\% reliability results in a total reliability of 98\,\%, c.f., \cref{fig:SimpleIntro}. When offers are stacked horizontally, the total reliability is thus higher than that of any individual offer. 

The vertical stacking of offers has the contrary effect: it decreases the joint reliability while it increases the volume, as it considers the sum of the offered volumes. For example, the vertical combination of 95\,\% and 95\,\% reliability results in a total reliability of 90.25\,\%. This implies that each vertical procurement block must have a higher reliability than the system reliability $\Phi^s$. When offers are stacked vertically, the total reliability is thus lower than the smallest reliability of its components.

\subsection{Probabilistic Formulation}
As illustrated in \cref{fig:offerstacking}, offers can be stacked horizontally and vertically. However, since only the volume can be decomposed, while the reliability cannot be decomposed, we propose the following sequence for the stacking of offers.
\begin{enumerate}
  \item[(a)] \textbf{Offers $i$ can be stacked horizontally} into procurement blocks $b=\{b_1,\dots,b_k\}$ to achieve the target reliability of each block. The procurement block volume $q_b$ is then limited by the smallest accepted offer $\{q_{i,b}\}$ in \cref{eq:relstacking1}, while the procurement block reliability $\phi_b$ is obtained with equation \cref{eq:relstacking2}. 
  \begin{subequations}
    \begingroup
    \allowdisplaybreaks
    \begin{align}
  	  & q_b = \underset{i}{\text{min}}\{q_{i,b}\} \quad\quad\quad \forall i,b \label{eq:relstacking1}\\
	  & 1-\phi_b = {\textstyle\prod_{i=i_1}^{i_n}} (1-R_{i} z_{i,b}) \label{eq:relstacking2}
    \end{align}
    \endgroup
    \label{eq:relstacking}
  \end{subequations}
  The binary variable $z_{i,b}$ is $1$ if offer $i$ is (partially) accepted in procurement block $b$ and $0$ otherwise. Horizontally, volumes do not sum, but reliability increases with every additional offer. \emph{When we rely on any of $\omega$ offers with the same volume being available, the failure of up to $\omega-1$ offers still leaves sufficient reserve volume.}
  \item[(b)] \textbf{Procurement blocks $b$ can be stacked vertically} to reach the volume $Q^s$ required by the SO. Vertically, volumes sum \cref{eq:volstacking1}, at the cost of lowered reliability \cref{eq:volstacking2}. 
  \begin{subequations}
    \begingroup
    \allowdisplaybreaks
    \begin{align}
  	  & {\textstyle\sum_{b=b_1}^{b_k}} q_b \ge Q^{s} \label{eq:volstacking1}\\
	  & {\textstyle\prod_{b=b_1}^{b_k}}\phi_b \ge \Phi^{s} \label{eq:volstacking2}
    \end{align}
    \endgroup
    \label{eq:volstacking}
  \end{subequations}
  Note, that the total reliability is lower than the smallest procurement block reliability, i.e., $\Phi^{s} \le \phi_b \; \forall b$. \emph{When we rely on the volume of $\beta$ blocks available at the same time, the failure of only block leads to insufficient volume.}
\end{enumerate}

\begin{figure}[t]
\centering
\resizebox{\columnwidth}{!}{\input{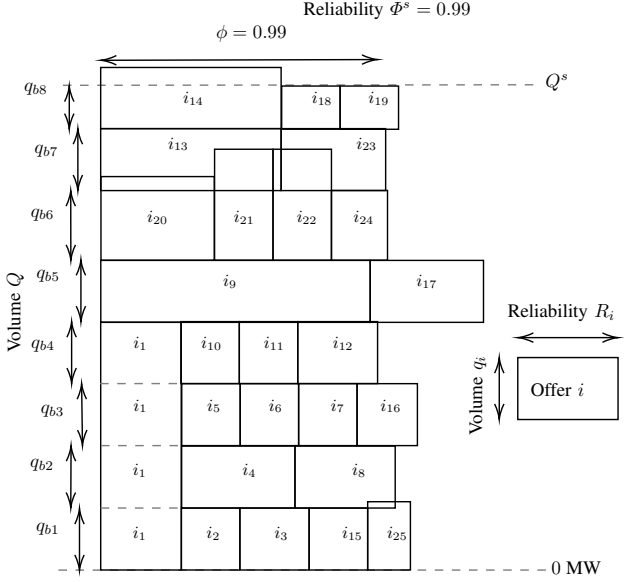}}
\caption{Illustration of stacking in procurement blocks based on an algorithm. The dimension and indexes of offer blocks is the same as in \cref{fig:offerstacking}.}
\label{fig:reliabilitystacking}
\vspace{-0.2cm}
\end{figure}
The assumption that a \emph{bid can be split by volume} seems more realistic than the assumption that a \emph{bid can be split into two parts with different reliabilities}, as in the latter case, the supplier would have submitted two different offers with different reliabilities to start with.
The horizontal combination of multiple offers $i$ into procurement blocks $b$ is illustrated in \cref{fig:reliabilitystacking} as the result of an algorithm. Offer $i_1$ has a large volume and is split into four different procurement blocks.

\subsection{Market Clearing}
The market clearing with reliability-aware reserve offers is formulated as a procurement cost minimization problem
%\newpage
%\small
\begin{subequations}
  \begingroup
  \allowdisplaybreaks
  \begin{align}
  	& \underset{q_{i,b},q_b,\phi_b,z_{i,b}}{\text{min.}} && 
	 \sum_{b} \sum_{i} q_{i,b} P_{i} & \label{eq:oA} \\
	& \text{s.t.} && q_b - q_{i,b} \le M (1-z_{i,b}) & \forall i,b \label{eq:cA1}\\
	& && 1-\phi_b = \prod_{i} (1-R_{i} z_{i,b}) \;& \forall b \label{eq:cA2}\\
	& && Q^{s} \le \sum_{b} q_b & \label{eq:cA3}\\
	& && \Phi^{s} \le \prod_{b}\phi_b & \label{eq:cA4}\\
	& && \sum_{b} q_{i,b} \le V_{i} \;& \forall i \label{eq:cA5}\\
	& && z_{i,b} \in \{0,1\} \;& \forall i,b \label{eq:cA6}\\
	& && q_{i,b}, q_b \ge 0 \;& \forall i,b \label{eq:cA7} \\	
	& && q_b \ge \underline{B} \;& \forall b \label{eq:cA8}
  \end{align}
  \endgroup
  \label{eq:problemA}
\end{subequations}
%\normalsize
where the SO's objective is to minimise the total cost paid to RRPs in \cref{eq:oA}. 
Constraints \cref{eq:cA1,eq:cA2} define the procurement block volume and reliability as in \cref{eq:relstacking}, where $M$ is a sufficiently large parameter. Constraints \cref{eq:cA3,eq:cA4} define the stacking of procurement block volumes as in \cref{eq:volstacking}.
Constraint \cref{eq:cA5} limits the procured quantity with the offered volume, and naturally offers must be positive \cref{eq:cA7}.
Constraint \cref{eq:cA8} is optionally added to reduce the solution time, where the SO may choose the minimum procurement block quantity $\underline{B}$.
Note that there are two motivations for selecting a large minimum procurement block quantity; (i) reliability and (ii) solution time.
\begin{itemize}
    \item[(i)] The lower the number $k$ of vertically stacked procurement blocks the higher the reliability, all else being equal. This is due to fewer factors in constraint \cref{eq:cA4}.
    \item[(ii)] The higher the minimum procurement block quantity $\underline{B}$ the faster the solution time.
\end{itemize}
Set $\mathcal{B}$ contains all procurement blocks $b=\{b_1,\dots,b_k\}$. The number of procurement blocks $b_k$ is generally not fixed. A simple method can be to set $b_k={ceil}\{\tfrac{Q^s}{\underline{S}}\}$ where $\underline{S}$ is the minimum bid size.

Note that the procured quantity $q_{i,b}$ of offer $i$ may be distributed in one, several or all procurement blocks. The set $\mathcal{I}$ contains all offers $i,\dots,i_n$, and thus the binary variable $z_{i,b}$ indicates which of the offers $i$ are (fully or partially) accepted in procurement block $b$.

\section{Mathematical Problem Reformulations} \label{sec:problem}
Constraints \cref{eq:cA2,eq:cA4} include bilinear terms which render the problem non-convex. This section first presents an equivalent reformulation which eliminates the bilinear terms of \cref{eq:problemA}. Second, it lays out a MILP approximation. Finally, it sketches further simplification approaches.
\subsection{Equivalent Problem Reformulation}
The use of logarithmic law $\ln(\prod_i x_i) = \sum_i \ln(x_i)$ allows us to reformulate \cref{eq:cA2} and \cref{eq:cA4} as
%\small
\begin{subequations}
  \begingroup
  \allowdisplaybreaks
  \begin{align}
    & \ln(1-\phi_b) = \sum_{i} \ln(1-R_{i} z_{i,b}) \;\forall b \label{eq:cA22}\\
    & \ln(\Phi^{s}) \le \sum_{b} \ln(\phi_b) \label{eq:cA42}
  \end{align}
  \endgroup
  \label{eq:problemC24}
\end{subequations}
%\normalsize
where inequality \cref{eq:cA42} describes a convex exponential cone.
Furthermore, we note that for $R_i\in[0,1)$ and $z_{i,b}\in\{0,1\}$ it holds that
%\small
\begin{equation}
    \ln(1-R_{i} z_{i,b}) = z_{i,b} \ln(1-R_{i}) \;\;\forall i,b. \label{eq:cA23}
\end{equation}
%\normalsize
Consequently, the right hand side of \cref{eq:cA22} can be reformulated by exploiting \cref{eq:cA23} which leaves only parameters inside the logarithm.
The problem formulation then reads
%\small
\begin{subequations}
  \begingroup
  \allowdisplaybreaks
  \begin{align}
  	& \underset{q_{i,b},q_b,\phi_b,z_{i,b}}{\text{min.}} && 
	 \cref{eq:oA} & \notag \\
	& \text{s.t.} && \text{\cref{eq:cA1,eq:cA3,eq:cA5,eq:cA6,eq:cA7,eq:cA8}} & \notag\\
	& && \ln(1-\phi_b) = \sum_{i} z_{i,b} \ln(1-R_{i}) & \;\forall b \label{eq:cC2}\\
	& && \ln(\Phi^{s}) \le \sum_{b}\ln(\phi_b) & & \label{eq:cC4}
  \end{align}
  \endgroup
  \label{eq:problemC}
\end{subequations}
We refer to the reformulated MINLP \cref{eq:problemC} as rMINLP. The problem is still non-convex due to equation \cref{eq:cC2}. However, the non-convexity is reduced with respect to \cref{eq:problemA}, since all bilinear terms have been eliminated from the formulation.

\subsection{Problem Relaxation \& Linearization}
This subsection introduces further assumptions that simplify the general market framework to a more practical and tractable one. This allows to reformulate the problem as a MILP which can then be solved to global optimality with branch and bound algorithms.

Constraint \cref{eq:cC4} includes the non-linear term $\ln(1-\phi_b)$ which can be relaxed with the assumption that each block must maintain a pre-specified reliability level $\Psi_b$ where $\Psi_b \le \phi_b \,\forall b$. The reliability level is computed offline by the SO and may, for instance, be uniformly distributed among all blocks, according to $\Psi_b =(\Phi^s)^{\frac{1}{b_k}} \,\forall b$. In fact, the most efficient way is to set $\Psi=\Psi_b$ constant $\forall b$, since $\Phi^s \le \underset{b}{\text{min}}\{\Psi_b\}$. 
The market clearing problem can then be approximated by
%\small
\begin{subequations}
  \begingroup
  \allowdisplaybreaks
  \begin{align}
  	& \underset{q_{i,b},q_b,z_{i,b}}{\text{min.}} && 
	 \cref{eq:oA} & \notag \\
	& \text{s.t.} && \text{\cref{eq:cA1,eq:cA3,eq:cA5,eq:cA6,eq:cA7}} & \notag \\
	& && \ln(1-\Psi) \ge \sum_{i} z_{i,b} \ln(1-R_{i}) \;& \forall b \label{eq:cD2}
  \end{align}
  \endgroup
  \label{eq:problemD}
\end{subequations}
where $\Psi$ is a parameter that replaces the variable $\phi_b$.
Constraint \cref{eq:cD2} is the linearised approximation of \cref{eq:cA2}. 

\subsection{Further Simplification}
Additionally, the SO may want to dictate a (minimum) offer reliability for each procurement block $b$ individually, such that $R_i\ge R_b\,\forall i$. We can further assume equality with uniform offer reliability $R_i z_{i,b}=R_b$ within each procurement block.
For uniform $R_i$, the minimum number of required offers per procurement block, in order to achieve a procurement block reliability $\Psi_b$ is 
\begin{equation}
    \sum_{i} z_{i,b} \ge \frac{\ln(1-\Psi_b)}{\ln(1-R_b)} \; \forall b \label{eq:cE2}  
\end{equation}
We can then write the market clearing problem as 
\begin{subequations}
  \begingroup
  \allowdisplaybreaks
  \begin{align}
	& \underset{q_{i,b},q_b,z_{i,b}}{\text{min.}} && 
	 \cref{eq:oA} & \notag \\
	& \text{s.t.} && \text{\cref{eq:cA1,eq:cA3,eq:cA5,eq:cA6,eq:cA7,eq:cA8,eq:cE2}} & \notag \\
	& && q_{i,b} R_{b} - \Psi_b \ge z_{i,b}-1 \;&\forall i,b \label{eq:cE3}
	\end{align}
  \endgroup
  \label{eq:problemE}
\end{subequations}
where \cref{eq:cE3} ensures non-zero volumes for accepted offers.

\section{Numerical Investigation} \label{sec:Case}
In this section, we first provide a small case study to illustrate the different formulations. We then compare the formulations to the reliability-unaware benchmark, present a large case study, and conduct sensitivity analysis with respect to block size and cost assumption.

\subsection{Small Case Study: Impact of Problem Reformulation \& Relaxation}
In order to compare the impact of different problem formulations, we first consult a small exemplary case study with 6 offers as in \cref{tab:Input1}, with $\underline{B}=$20\,MW and $Q^{s}=$40\,MW with reliability $\Phi^{s}=$99.95\,\%. 
The motivation for such a high reliability requirement is grounded in the massive cost to society and equipment in case of a power system outage.
For simplicity, the offer prices are assumed to increase linearly with reliability according to $P^\text{lin}=\alpha R$ where $\alpha=$100\,\textdollar.
\begin{table}[t]
  \centering
  \caption{Reliability-Aware Reserve Offers}
    %\resizebox{0.75\columnwidth}{!}{%
    \setlength{\tabcolsep}{4pt}
    \begin{tabular}{c rrr}
    \hline
    Offer  & Volume  & Reliability & Price       \\
    \hline
    1       & 40 MW  &       80 \% & 80 \textdollar/MW \\
    2       & 30 MW  &       90 \% & 90 \textdollar/MW \\
    3       & 30 MW  &       95 \% & 95 \textdollar/MW \\
    4       & 30 MW  &       98 \% & 98 \textdollar/MW \\
    5       & 20 MW  &       99 \% & 99 \textdollar/MW \\
    5       & 20 MW  &       99 \% & 99 \textdollar/MW \\
    \hline 
    \end{tabular} %}
  \label{tab:Input1} 
\vspace{-0.2cm}
\end{table}
None of the offers alone can satisfy the required system reliability, but the probabilistic clearing can. The results of applying the MINLP \cref{eq:problemA} with bilinear terms, reformulated rMINLP \cref{eq:problemC}, and relaxed MILP \cref{eq:problemD} formulations are summarised in \cref{tab:Results01}.
\begin{table}[t]
  \centering
  \caption{Impact of Problem Relaxation}
    %\resizebox{0.75\columnwidth}{!}{%
    \setlength{\tabcolsep}{4pt}
    \begin{tabular}{l rrrr}
    \hline
    Problem &                   Cost & Volume & Reliability $\Phi$ & Time \\
    \hline
    Unaware                   &  3,960 \textdollar &  40 MW    & $^*$98.010 \% &  $^*$(infeas.)\\
    MINLP \cref{eq:problemA}  &  9,320 \textdollar & 100 MW    & 99.950 \% & 265 ms\\
    rMINLP \cref{eq:problemC} &  9,320 \textdollar & 100 MW    & 99.950 \% & 203 ms\\
    MILP \cref{eq:problemD}   &  9,420 \textdollar & 100 MW    & 99.975 \% &  31 ms\\
    \hline 
    \end{tabular} %}
  \label{tab:Results01} 
\vspace{-0.2cm}
\end{table}

Both MINLP and rMINLP result in the same market outcome which satisfies \cref{eq:cA4,eq:cC4} with equality. The total cost and volume are the same, while the solution time is faster for the rMINLP.
The MILP approximates the solution of the MINLP with a gap of 100\,\textdollar \;(1.1\,\%). However, since the MILP approximation is more conservative, the overall system reliability is higher. Furthermore, the solution time of the MILP is faster compared to the MINLP and rMINLP.

\subsection{Benchmark: Reliability-Unaware Clearing}
For comparison, the reliability-unaware benchmark is listed in \cref{tab:Results01} where two offers of $20$\,MW and $99$\,\% reliability would be cleared. In this case, the total reliability of $99.84$\,\% is below the system requirement $\Phi^s$. Thus, the reliability unaware clearing would not yield a feasible solution that can satisfy the required system reliability. This simple example illustrates one of the shortcomings of today's reserve markets which cannot capture the full uncertainty of reserve providers.

\subsection{Large Case Study: National Reserve Market} \label{sec:LargeCase}
We use bids with reliability resolution $R_i=\{.01, .02, \dots, .99\}$ and volume $V_i=500$\,MW $\forall i$, while the SO's requirement is $Q^{s}=500$\,MW at $\Phi^{s}=0.9995$. Furthermore, we divide the procurement into 5 blocks of 100\,MW each. The results are summarised in \cref{tab:Results02}.
\begin{table}[t]
  \centering
  \caption{Impact of Problem Relaxation with 5 blocks of 100 MW}
    %\resizebox{0.75\columnwidth}{!}{%
    \setlength{\tabcolsep}{4pt}
    \begin{tabular}{l rrrr}
    \hline
    Problem                  &                 Cost & Volume & Reliability $\Phi$ & Time \\
    \hline
    Unaware   &   50,000\;\textdollar &      500\;MW & $^{*}$99.000\;\% & $^*$(infeas.)\\    %Deterministic & 200,000\;\textdollar &      500\;MW & 100.000\;\% & 0.015\;s\\
    MINLP \cref{eq:problemA} &  207,100\;\textdollar & 2,400\;MW & 99.995\;\% & 1003.230\;s \\
    rMINLP \cref{eq:problemC} & 146,700\;\textdollar & 1,500\;MW & 99.995\;\% & 3605.700\;s \\
    MILP \cref{eq:problemD} &   147,000\;\textdollar & 1,500\;MW & 99.995\;\% & 0.024\;s\\
    \hline 
    \end{tabular} %}
  \label{tab:Results02} 
\vspace{-0.2cm}
\end{table}

Again, the reliability unaware clearing cannot achieve sufficiently high reliability (99.0\;\%) with the available offers. However, it would result in the lowest total procurement cost.
In this larger case study, the MINLP \cref{eq:problemA} does not converge to the global optimum. The equivalent rMINLP \cref{eq:problemC} yields a lower objective. The MILP finds a solution close to the one of the rMINLP with a gap of 300\,\textdollar, i.e., 0.002\,\%. 
Furthermore, the solution time becomes crucial in this larger case study. Only the MILP can clear the market in comparable time scales as the state of the art reliability-unaware clearing. The MINLP and rMINLP need more than 15 minutes which may be critical to large power systems and reserve markets.

Note that, in comparison to \cref{tab:Results01}, here, the rMINLP solves slower than the MINLP. The rMINLP, however, converges to a 30\,\% lower procurement cost. In this setup, we have selected the default convergence criteria of GAMS with the DIPLEX solver. Since the MINLP is highly non-convex and the convergence strongly depends on the problem size, the solver stopped after 1003.23\;ms in our case study, and is still further from the optimum than the relaxed MILP.

\subsection{Sensitivity to Procurement Block Size}
For national level reserve markets in the range of hundreds of MWs further practical challenges arise. The SO needs to define the number (or size) of procurement blocks which are parameters in the problem formulation. On the one hand, the SO would want to keep a low number of blocks in order to increase overall reliability, c.f., equation \cref{eq:volstacking2}. On the other hand, the SO would want to keep a low block volume in order to increase liquidity in the market from smaller RRPs. 

Here, we use the MILP to study the effect of different block sizes on the cost, total procured volume, block reliability and solution time in \cref{tab:Results03}. We assume constant liquidity among all cases, i.e., the set of offers detailed in \cref{sec:LargeCase}.
\begin{table}[t]
  \centering
\caption{Impact of minimum block size $\underline{B}$ in MILP}
    %\resizebox{0.75\columnwidth}{!}{%
    \setlength{\tabcolsep}{4pt}
    \begin{tabular}{r rrrrr}
    \hline
    Block Size & Blocks &                Cost & Volume   & Block Reliability & Time \\
    \hline
    500\;MW     &      1 & 145,000\;\textdollar & 1,500\;MW & 99.99500\;\% & 16\;ms\\
    250\;MW     &      2 & 147,000\;\textdollar & 1,500\;MW & 99.99750\;\% & 20\;ms\\
    100\;MW     &      5 & 147,000\;\textdollar & 1,500\;MW & 99.99900\;\% & 24\;ms\\
     50\;MW     &     10 & 162,000\;\textdollar & 2,000\;MW & 99.99950\;\% & 32\;ms\\
     25\;MW     &     20 & 188,000\;\textdollar & 2,000\;MW & 99.99970\;\% & 38\;ms\\
     10\;MW     &     50 & 195,000\;\textdollar & 2,000\;MW & 99.99990\;\% & 78\;ms\\
      5\;MW     &    100 & 195,000\;\textdollar & 2,000\;MW & 99.99995\;\%& 141\;ms\\
      2\;MW     &    250 & 210,000\;\textdollar & 2,500\;MW & 99.99998\;\%& 312\;ms\\
      1\;MW     &    500 & 236,000\;\textdollar & 2,500\;MW & 99.99999\;\%& 760\;ms\\
    \hline 
    \end{tabular} %}
  \label{tab:Results03} 
\vspace{-0.2cm}
\end{table}
As the block size decreases, more blocks are required, which enforces increasingly higher reliability $\Psi$ on each block. This leads to both an increase in procurement cost and volume. We also observe an increase in solution time which is, however, not considered critical for the practical time scales of reserve markets.

\subsection{Sensitivity to Cost Assumption}
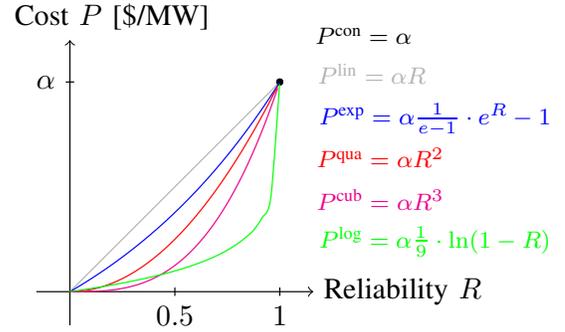
\begin{figure}[t]
\centering
\resizebox{0.85\columnwidth}{!}{\begin{tikzpicture}
  \draw[->] (-0.4, 0) -- (2.9, 0) node[right] {Reliability $R$};
  \draw[->] (0, -0.4) -- (0, 3.0) node[above, xshift=0.5cm] {Cost $P$ [\textdollar/MW]};
  %\draw[  ] (-0.4, 0) -- (2.9, 0) node[above, xshift=0.4cm, yshift=-0.7cm] {Reliability};
  %\draw[scale=0.5, domain=-3:3, smooth, variable=\x, blue] plot ({\x}, {\x*\x});
  %\draw[scale=1.0, domain=0:0.99, smooth, variable=\x, blue] plot ({\x}, {(1/100)/(1-\x)-(1/100)} );
  %\draw[scale=1.5, domain=0:0.99, smooth, variable=\x, blue] plot ({\x}, {-(1/9.2103)*ln(1-\x)} );

  %\draw[scale=1.0, domain=0:0.99, smooth, variable=\x, green] plot ({\x}, {e^(0.7*\x)-1} );
  \draw[scale=2.5, domain=1:1, smooth, mark=*, mark options={solid, black}, mark size=0.4pt, variable=\x, black] plot ({\x}, {(1*\x)} );
    \node at (3.5,3.05) {\footnotesize\color{black} $P^\text{con}=\alpha$};
  \draw[scale=2.5, domain=0:0.99, smooth, variable=\x, black!30!white] plot ({\x}, {(\x)} );
    \node at (3.6,2.6) {\footnotesize\color{black!30!white} $P^\text{lin}=\alpha R$};
  \draw[scale=2.5, domain=0:0.999, smooth, variable=\x, blue] plot ({\x}, {(1/(exp(1)-1))*(exp(\x)-1)} );
    \node at (4.35,2.05) {\footnotesize\color{blue} $P^\text{exp}=\alpha \frac{1}{e-1}\cdot e^{R}-1$};
  %\draw[scale=0.5, domain=-8:8, smooth, variable=\x, blue] plot ({\x}, {1/\x});
  \draw[scale=2.5, domain=0:0.99, smooth, variable=\x, red] plot ({\x}, {(\x)*(\x)} );
    \node at (3.7,1.6) {\footnotesize\color{red} $P^\text{qua}=\alpha R^2$};
  \draw[scale=2.5, domain=0:0.99, smooth, variable=\x, magenta] plot ({\x}, {(\x)*(\x)*(\x)} );
    \node at (3.7,1.1) {\footnotesize\color{magenta} $P^\text{cub}=\alpha R^3$};
  \draw[scale=2.5, domain=0:0.9992, smooth, variable=\x, green] plot ({\x}, {-(1/7)*ln(1-\x)} );
    \node at (4.35,0.6) {\footnotesize\color{green} $P^\text{log}=\alpha \frac{1}{9}\cdot\ln(1-R)$};
    
  \draw[-, black] (1.25, -0.05) -- (1.25, .05) node[below, yshift=-0.1cm] {$0.5$};
  \draw[-, black] (2.5, -0.05) -- (2.5, .05) node[below, yshift=-0.1cm] {$1$};
  \draw[-, black] (-0.05, 2.5) -- (.05, 2.5) node[left, xshift=-0.1cm] {$\alpha$};
  %\draw[-, black] (-0.05, 5) -- (.05, 5) node[left, xshift=-0.1cm] {$4\alpha$};
    
\end{tikzpicture}}
\caption{Cost over reliability under different assumptions.}
\label{fig:costcurve}
\vspace{-0.2cm}
\end{figure}

A linear relationship of cost and reliability is rather conservative, considering the increased volume from a strategic ELR perspective. We therefore study the impact of a range of cost functions that are illustrated in \cref{fig:costcurve}. The linear, exponential, quadratic, and cubic price functions all intersect at reliability 0 and 1.
Note that an offer with $50$\,\% reliability corresponds to a coin flip, which justifies the assumption of cubic or logarithmic price function.
We use the same offers ans SO requirement as in \cref{sec:LargeCase}. The results from using different price assumptions are listed in \cref{tab:Results04}.
\begin{table}[t]
  \centering
  \caption{Reliability-Aware Reserve Offers}
    %\resizebox{0.85\columnwidth}{!}{%
    \setlength{\tabcolsep}{4pt}
    \begin{tabular}{l rrrr}
    \hline
    Type & Cost [$\alpha$\textdollar]& Volume [MW]& Reliability \\
    \hline
    Unaware       & 500.0    &   500   & 99.840\;\%  \\
    Linear        & 949.9    & 1,000   & 99.995\;\%  \\
    Exponential   & 895.1    & 1,000   & 99.995\;\%  \\
    Quadratic     & 799.9    & 4,500   & 99.995\;\%  \\
    Cubic         & 611.4    & 4,500   & 99.995\;\%  \\
    Logarithmic   & 625.0    & 2,300   & 99.995\;\%  \\
    \hline 
    \end{tabular} %}
  \label{tab:Results04} 
\vspace{-0.2cm}
\end{table}

The procurement cost is lowest for the cubic and logarithmic price functions, where price levels are generally lower. 
The total procured volume is highest for quadratic and cubic price functions. This is due to the low prices for low reliability ($\le 0.5$) bids, where additional volume needs to be aggregated into a procurement block to reach the same block reliability $\Psi$.
This observation in \cref{tab:Results04} underlines that the behavior of procurement volume, cost and reliability are decoupled.

\section{Discussion \& Possible Extensions} \label{sec:Disc}
This section discusses methods to mitigate the dependence of reserve offers, fairness issues and offering strategy.
\subsection{Correlation of Reliability from Renewable Energy Sources}
Constraints \cref{eq:cA2,eq:cA4} are based on the assumption that the reliability of reserve offers is independent of each other. However, due to shared weather dependence, this assumption does not hold in practice for renewable energy sources. Hence, one may need different versions of equations \cref{eq:relstacking2,eq:volstacking2} to account for various dependency models, and possibly a learning approach for that dependence.
Here, we lay out two approaches to mitigate the correlation of renewable energy sources which share - at least in part - the same uncertainty source. Note, however, that both approaches can only mitigate this dependence, but not eliminate it.

\subsubsection{Restrictions on source type in procurement block}
We assume that weather dependence is only shared between offers that origin from the same renewable energy source $s\in\mathcal{S}$. Set $\mathcal{S}$ includes different reserve sources (wind, solar, etc.) which is an additional attribute of reserve offers.
The SO can then decide to only allow bids from 'sufficiently different' renewable reserve sources, where the definition of 'sufficiently different' depends on the SO's classification and risk-aversion.
The problem can be formulated as
\begin{subequations}
  \begingroup
  \allowdisplaybreaks
  \begin{align}
  	& \underset{q_{i,b},q_b,z_{i,b}}{\text{min.}} && 
	 \sum_{b} \sum_{i} \sum_{s} q_{i,b} P_{i} \label{eq:oG} \\
	& \text{s.t.} && \text{\cref{eq:cA5,eq:cA6,eq:cA7,eq:cA8}} & \notag \\	
    & && q_b - \sum_s U_{s,i} q_{i,b} \le M (1-z_{i,b}) \; \forall i,b \label{eq:cG1}\\
	& && \ln(1\!-\!\Psi) \!\ge\! \sum_{i} \sum_s U_{s,i} z_{i,b} \ln(1\!-\!R_{i}) \; \forall b \label{eq:cG2}\\
	& && \ln(\Phi^{s}) \le \sum_{b}\ln(\phi_b) & & \label{eq:cG4}\\
	& && \sum_{b} q_{i,b} \le V_{i} \; \forall s,i \label{eq:cG5}\\
	& && \sum_i U_{s,i} z_{i,b} \le 1 \; \forall s,b \label{eq:cG9}
  \end{align}
  \endgroup
  \label{eq:problemG}
\end{subequations}
where the RRP submits a binary source indicator $U_{s,i}$ that is $1$ for exactly one $s\in\mathcal{S}\,\forall i$. 
This indicator ensures in constraint \cref{eq:cG9} that only offers from sufficiently different reserve offer sources are accepted in each procurement block. In other words, maximum one wind offer, and maximum one solar offer is allowed in each procurement block. In that way, the independence assumption holds for \cref{eq:relstacking2}. However, the independence of vertically stacked blocks \cref{eq:volstacking2} is not guaranteed. Furthermore, the limited number of sources $s$ would quickly reduce the total reserve volume that can be provided.

\subsubsection{Accounting for Cross-correlation} 
If the reliability dependency $\varrho_{i,j}$ of offers $i$ and $j$ is known, we can gather this information in a cross-correlation matrix $\Gamma$.
This information can then be included as parameters in problem \cref{eq:problemC} as
\begin{subequations}
  \begingroup
  \allowdisplaybreaks
  \begin{align}
    & \underset{q_{i,b},q_b,z_{i,b}}{\text{min.}} && 
	 \cref{eq:oA} & \notag \\
	& \text{s.t.} && \text{\cref{eq:cA1,eq:cA3,eq:cA5,eq:cA6,eq:cA7,eq:cA8}} & \notag \\
	& && \ln(1\!-\!\Psi) \ge \sum_{i} \prod_j(\varrho_{i,j}) z_{i,b} \ln(1\!-\!R_{i}) \; \forall b \label{eq:cF2}\\
	& && \ln(\Phi^{s}) \le \sum_{b} ( \ln(\phi_b) \sum_{i} \prod_j(\varrho_{i,j}) z_{i,b} ) \label{eq:cF4}
  \end{align}
  \endgroup
  \label{eq:problemF}
\end{subequations}

In practice, we would not know $\Gamma$ for several reasons. Past correlations cannot predict future correlation due to unique ambient conditions. However, we can approximate $\Gamma$ to some degree. Furthermore, this approximation of $\Gamma$ can be continuously improved using an online learning approach.

\subsection{Fairness}
Unfair allocation implies that sub-optimality is introduced in the market clearing, which is unevenly distributed among reserve providers. For example, one offer will get accepted although another correlated offer could have provided the reserve at equal or lower cost. In our future research we aim to establish analytical formulations and an upper bound on this sub-optimality gap.

Here, we assume transparency and honesty of RRPs about their reliability. In practice, a market mechanism must be established that incentivises truthful bidding with respect to reliability. For instance, the market operator may average the observed availability of offer $i$ over a long enough time horizon in order to compare against their stated reliability. If the deviation exceeds a certain threshold, penalties may be established to ensure truthful bidding.

\subsection{Offering Strategy}
Different cost assumptions were analysed in this paper, since there is no practical evidence from reliability-aware offers. In practice, RRPs would need to solve a separate optimization problem to divide their capacity into blocks of different volume and reliability, together with the cost. The most suitable RRP offering strategy will probably vary depending on the reserve source; in our future work we intend to further investigate this.

\section{Conclusions} \label{sec:Conc}
In this paper, we introduce the novel concept of reliability-aware probabilistic reserve procurement. We detail the cost minimization problem of selecting sufficient reserves while maintaining specified reliability criteria and demonstrate the cost efficacy of probabilistic reserve procurement. 
The proposed approach increases liquidity, lowers cost, and enables previously `unreliable' RRPs such as renewabke energy sources to offer even their uncertain capacity.
We compare the proposed approach to the state-of-the-art reliability-unaware market clearing in terms of overall cost, volume, and reliability.
We further introduce two approximations that reduce the solution time by 5 orders of magnitude while maintaining a good performance (0.002\,\% optimality gap for a large power system).

%\section*{Acknowledgment}

\bibliographystyle{IEEEtran}
\bibliography{ReserveClearing}

\end{document}